# Human-like Nonverbal Behavior with MetaHumans in Real-World Interaction Studies: An Architecture Using Generative Methods and Motion Capture

Oliver Chojnowski
*Cologne Cobots Lab*
*TH Köln – University of Applied Sciences*
50679 Cologne, Germany
oliver.chojnowski@th-koeln.de
0009-0003-9045-8786

Alexander Eberhard
*Cologne Cobots Lab*
*TH Köln – University of Applied Sciences*
50679 Cologne, Germany
alexander.eberhard@th-koeln.de
0009-0001-2527-9250

Michael Schiffmann
*Cologne Cobots Lab*
*TH Köln – University of Applied Sciences*
50679 Cologne, Germany
michael.schiffmann@th-koeln.de
0000-0002-7328-9859

Ana Müller
*Cologne Cobots Lab*
*TH Köln – University of Applied Sciences*
50679 Cologne, Germany
ana.mueller@th-koeln.de
0000-0002-4960-082X

Anja Richert
*Cologne Cobots Lab*
*TH Köln – University of Applied Sciences*
50679 Cologne, Germany
anja.richert@th-koeln.de
0000-0002-3940-3136

*Abstract*— Socially interactive agents are gaining prominence in domains like healthcare, education, and service contexts, particularly virtual agents due to their inherent scalability. To facilitate authentic interactions, these systems require verbal and nonverbal communication through e.g., facial expressions and gestures. While natural language processing technologies have rapidly advanced, incorporating human-like nonverbal behavior into real-world interaction contexts is crucial for enhancing the success of communication, yet this area remains underexplored. One barrier is creating autonomous systems with sophisticated conversational abilities that integrate human-like nonverbal behavior. This paper presents a distributed architecture using Epic Games' MetaHuman, combined with advanced conversational AI and camera-based user management, that supports methods like motion capture, handcrafted animation, and generative approaches for nonverbal behavior. We share insights into a system architecture designed to investigate nonverbal behavior in socially interactive agents, deployed in a three-week field study in the Deutsches Museum Bonn, showcasing its potential in realistic nonverbal behavior research.

## I. Introduction

Socially interactive agents (SIAs) aim to enable natural and intuitive human-machine interaction by incorporating human face-to-face communication modalities into machine-based systems. To interact with humans in a socially intelligent manner, nonverbal behaviors such as facial expressions and gestures play a crucial role [1]. Although significant progress has been made in advancing verbal communication technologies, including Speech-to-Text (STT), Text-to-Speech (TTS), Natural Language Understanding (NLU), and Large Language Models (LLMs), the generation and integration of meaningful nonverbal behaviors in SIAs remains challenging.

Motion capture technology allows the representation and transfer of human movement onto SIAs with high fidelity, but it lacks the flexibility to autonomously react to new situations, an essential capability for studying real social interactions. To enable SIAs to autonomously hold conversations and react nonverbally, generative approaches are necessary to provide the required flexibility. As described in [2] and [3], the ultimate goal of gesture generation is to enhance communication and interaction between humans and agents, which requires evaluating these methods in actual interactions. However, relying on multiple components, such as automatic speech recognition (ASR), STT, TTS, and Conversational AI (CAI), to realize verbal interactivity makes such interaction studies rare. Other works, such as [4], [5], also emphasize the importance of moving beyond laboratory settings to study social interactions 'in the wild'.

In this paper, we present an architecture that enables the use of Epic Games' (Cary, USA) MetaHuman as an SIA, including all the necessary components and functionalities for autonomous interactions. This includes camera-based user management, which detects users, a sophisticated CAI, and the ability to integrate generative approaches for nonverbal behavior like facial animations alongside motion-capture for e.g., gestures. The modular and flexible nature of the architecture facilitates the efficient setup of different experimental conditions, such as using only motion-captured or only generative nonverbal behavior.

To address the ever-present challenge of achieving low processing times in SIAs, as noted by [3], the system is implemented as a distributed client-server architecture primarily utilizing Representational State Transfer (REST) and WebSocket connections. This approach allows individual modules to run on separate consumer-grade hardware resources, thereby reducing reaction and processing times, while also lowering costs and barriers to use.

Building upon this architecture, we share our insights and observations gained from development and deployment in a three-week field study in a public museum. This work contributes to the research on SIAs by providing an architecture that facilitates the study of nonverbal behaviors in autonomous real-world interaction scenarios.

## II. RELATED WORK

Nonverbal behavior in SIAs encompasses various modalities, including facial expressions, gaze, and gestures [6]. The implementation of nonverbal behavior in SIAs can be categorized into two main approaches: rule-based methods, where predefined rules dictate behavior, and data-driven generative models, which are trained to associate speech with gestures from large datasets. These datasets are typically created using motion capture techniques to extract video data of humans speaking, along with the corresponding audio and text transcripts. The motion-captured data, when transferred to virtual agents or social robots, also serves as a baseline for evaluating generative models. This approach is systematically pursued in the GENEA Challenge, which aims to establish a standardized benchmark for gesture generation [7,8,9]. A comprehensive review detailing the evolution and state-of-the-art developments in co-speech gesture generation is provided in [2].

Given the different modes of nonverbal communication and their diverse communicative functions, research often focuses on individual components, such as facial expressions or gestures. For instance, facial animations have advanced significantly, particularly through motion capture and deep learning methods, as demonstrated by the Audio2Face model [10]. In contrast, gesture generation remains one of the most challenging aspects of nonverbal behavior. Although the methods reviewed in [2] achieve high levels of human-likeness, they often fall short in speech appropriateness, which means that gestures do not align consistently with the accompanying verbal communication. Some systems attempt to generate full-body nonverbal behavior, such as the EMAGE framework [11], which incorporates facial, local body, hand, and global movements. However, full-body solutions face the additional challenge of computational complexity, which can lead to longer inference times. This increased delay reduces the responsiveness essential for social interactions.

The evaluation of nonverbal behavior generation methods is predominantly conducted in controlled, standardized settings. These evaluations often involve analyzing videos of agents performing specific movements to assess human-likeness and appropriateness of the nonverbal behavior to the speech. Although effective for benchmarking models, such evaluations do not reflect real-world interactions. Of all studies exploring gesture generation in interactions with Embodied Conversational Agents (ECAs), the majority remain confined to laboratory settings, with limited investigation into real-world social contexts. In a review of evaluation practices for gesture generation in ECAs, only one of 23 studies was conducted in the wild [12].

Although an architecture integrating dialog systems for ECAs with gesture generation models has been suggested in [13], it focuses primarily on the initial incorporation of gesture generation within conversational frameworks. However, it does not fully address the critical requirements for seamless and autonomous social interactions in dynamic real-world environments. Current systems often lack robust user detection mechanisms and may not efficiently combine the strengths of NLU and LLMs. Our work aims to fill this gap by extending existing architectures with vital enhancements, such as camera-based user management to detect user presence, and a CAI system that leverages a hybrid approach of NLU and LLMs. In doing so, we offer a more comprehensive solution for studying and enabling socially interactive agents in real-world scenarios.

## III. SYSTEM ARCHITECTURE

To address the identified limitations in existing architectures, we propose a comprehensive system that integrates advanced components for autonomous interactions in dynamic settings. It integrates components like a MetaHuman for embodiment, a hybrid dialog system with LLMs and NLU, camera-based user management, and generative methods for nonverbal behavior.

### A. Distributed System Architecture

The system is implemented as a distributed client-server architecture which allows components such as the STT or the model for generating facial animations to operate independently on separate consumer-grade hardware, optimizing processing times and ensuring scalability. WebSocket, Google Remote Procedure Calls (gRPC), Unreal Engine's LiveLink and REST connections are utilized for low-latency communication between clients and servers hosting the computational modules. The system operates on consumer-grade NVIDIA RTX GPUs (RTX 4060, 4070 and 4090), ensuring accessibility and by supporting locally running models, enhances GDPR compliance, ensuring data remains within the system. The modular nature of the architecture allows for customization of conversation flow and interaction logic, supporting various experimental conditions and behaviors. Individual modules can be swapped to integrate locally running models or invoke cloud-based models via Application Programming Interface (API), providing flexibility for various deployment scenarios. This modularity also makes the architecture agile and easily adaptable to the rapid advancements in AI technology.

### B. MetaHuman

The MetaHuman serves as the actual embodiment of the SIA, integrating a state-based logic to enable dynamic, context-aware interactions. A state machine governs the MetaHuman's behavior, transitioning between four states: idle, listening, thinking, and talking, as visualized in Figure 1. Both the idle and thinking states involve the MetaHuman exhibiting subtle, lifelike micro-movements, thereby maintaining an engaging and calm presence while waiting for user input or processing information. When in the Listening state, the agent visually indicates attention by slightly nodding its head. During the Talking state, the MetaHuman delivers verbal responses accompanied by synchronized nonverbal behaviors, such as facial animations and gestures.

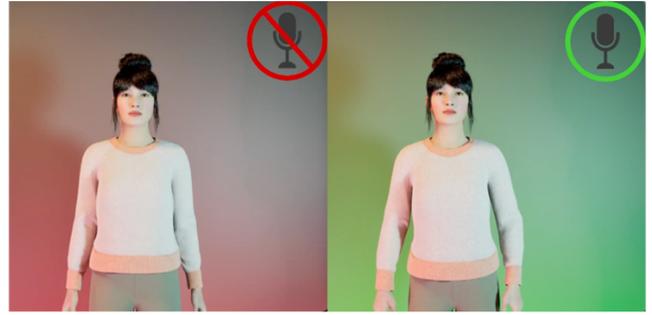

Figure 2: The MetaHuman as embodiement of the SIA. Background color and icon indicate if the agent is listening (right) or not (left)

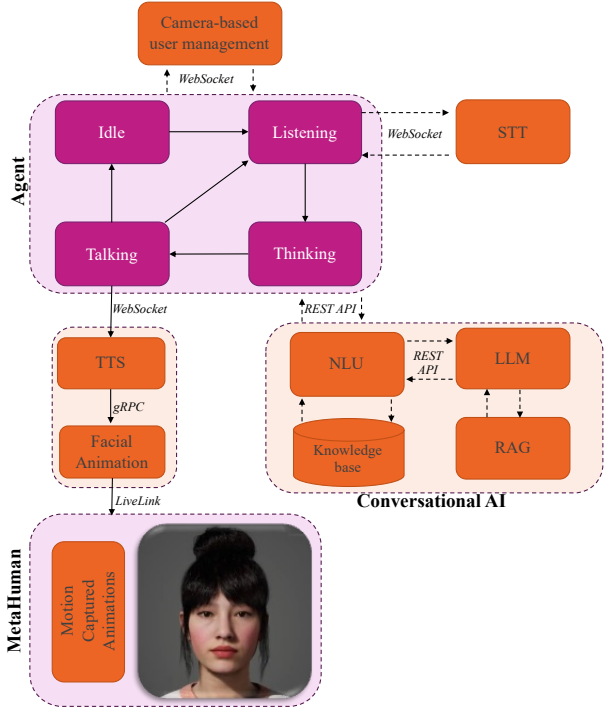

Figure 1: State-based architecture with the corresponding interfaces to the individual distributed models

To enhance the user experience and enable intuitive user-friendly interactions, the design incorporates clear visual indicators as displayed in Figure 2. The background color of the environment and symbolic elements dynamically adapt to the state of the MetaHuman. A red background with a "mute" symbol signifies that the agent is not listening, which is displayed in the idle, talking and thinking state. During the listening state a green background with a microphone icon signals that the user can speak, providing a clear prompt for interaction readiness.

*C. Hybrid Dialog System*

The dialog system in this architecture integrates NLU (Google, Dialogflow [14]) with an LLM (Meta AI, Llama 3.2 [15]) to manage conversations and generate responses as shown in Figure 1. The NLU module uses a predefined knowledge base to identify intents and match user queries to predefined topics, ensuring that the system can respond accurately and contextually. Simultaneously, the LLM operates as a fallback mechanism, enhancing the system's ability to handle open-ended queries or topics not covered by the knowledge base. We chose the Llama 3.2 3B model with EXL2 6.0 bits per weight quantization [16] for its ability to run locally on an RTX 4090 GPU, ensuring data privacy and GDPR compliance by avoiding reliance on cloud-based solutions. The selected response is then forwarded to the TTS module, which uses OpenAI's TTS service through a cloud API to generate speech output [17].

*D. Camera-based User-Management*

The camera-based user management system enables the agent to adjust its behavior based on user presence. It uses Google's MediaPipe EfficientDet-Lite0 model for user detection and distance estimation at 30 frames per second . The agent focuses on individuals within its interaction zone, approximating their distance by the relative size of the bounding box to the total image size to reduce computational demands. To robustly detect users of varying heights and positions, the system also checks if the midpoint of the bounding box exceeds a specified height, and that way can also detect, e.g., small children inside the interaction zone. Additionally, the system employs a frame-averaging mechanism that changes the agent's state only after consistent changes are detected over multiple frames, thereby enhancing interaction reliability and robustness by mitigating the impact of erroneous detections. The user management sends an event when users enter or leave. It also sends the number of people that are detected to adjust the answers of the CAI for individual users and groups, enabling the SIA to adapt its behavior in group interactions [19].

*E. Natural Language Interface*

The natural language interface integrates STT and TTS to enable fluid verbal interaction with the MetaHuman. For STT, the system employs a medium-sized OpenAI (San Francisco, USA) Whisper model locally running, ensuring robust and fast transcription of user utterances [20]. The interaction begins when the camera-based user management detects a user and sends a command, signaling the system to activate the ASR process.

The system listens until the user stops speaking for a predefined, adjustable duration (e.g., 700 milliseconds). Once this threshold is reached, the captured utterance is sent to the CAI, which determines whether the response should come from the predefined knowledge base or be generated dynamically by the LLM. The selected response is then forwarded to the TTS module, which utilizes OpenAI's TTS service via a cloud API to generate speech output, using the voice 'Nova' [17]. This Model was chosen because of the high human-likeness of the speech output.

*F. Real-Time Generation of Facial Animations*

The resulting audio file from the TTS is sent to Audio2Face, which processes it to create real-time facial animations for the MetaHuman. These animations are streamed directly to the

MetaHuman's face via Unreal Engine's LiveLink connection, ensuring that the visual expressions align seamlessly with the spoken response. When the animation streaming concludes, indicating that the MetaHuman has finished speaking, a command is issued. This signal allows the system to transition the MetaHuman back into the Listening state, ready to process new user inputs.

## IV. Hybrid Generative and Motion Captured Nonverbal Behaviour

### A. Motion Capture Methodology

The hybrid approach focuses on capturing body movements via motion capture while excluding facial expressions, as these are generated separately by Audio2Face. The motion capture for nonverbal behaviors is achieved through a camera-based method using Dollars MoCap (Taipei, Taiwan) proprietary Dollars DEEP Software. Postprocessing is required to refine the captured body movements, ensuring they are free of artefacts and transition smoothly from and back to idle mode. Once processed, the movements are tagged with an identifier that is stored with the corresponding answer in the knowledge base, enabling them to be referenced and applied during the interaction. This setup ensures that the pre-recorded nonverbal behaviors are robustly called from the CAI.

### B. Implementation

Motion capture is employed for frequently recurring speech acts derived from the knowledge base. These include common phrases like greetings, farewells, follow-up queries such as "How can I help you now?" or self-introductions. Each of these actions are captured and saved as assets within Unreal Engine, where they are associated with unique identifiers that are referenced to a specific answer from the knowledge base. When the corresponding speech act is triggered, the appropriate motion capture asset is selected based on its ID, and displayed on the MetaHuman. During non-frequently occurring speech acts the agent displays only generated facial animations using Audio2Face.

## V. Discussion

Our experience with deploying the proposed architecture for SIAs in a real-world setting has underscored several key insights, with the necessity of on-site testing and fine-tuning emerging as most important to ensure user engagement and satisfaction. Each component of the system, such as STT, TTS or the camera-based user management, required specific adjustments to factors such as ambient noise, lighting conditions, and user demographics. While each deployment demands tailored adjustments to harmonize with local nuances, the general architecture proved adaptable to the real-world scenario. This adaptability suggests that with sufficient context-specific fine-tuning, the system can be effectively transferred to a variety of applications, highlighting the architecture's generalizability across various settings.

We observed that system delays, due to insufficient computational power or prolonged API calls, due to network latency, can negatively impact the user experience and reduce the effectiveness of nonverbal communication, thus making the study of nonverbal behavior difficult.

A further observation relates to maintaining coherence in interactions between the NLU and LLM. In real-world applications, discrepancies frequently arose between the LLM and the knowledge base, resulting in contradictory responses that could confuse users. Moreover, the LLM can overlook the agent's embodiment, occasionally producing responses misaligned with the agent's physical presence. This reinforces the need for meticulous prompt engineering to align LLM responses with the knowledge base and the agent's inherent personality. Lastly, our findings underscore the need to meet user expectations regarding high-quality TTS and the specificity of system responses. Users naturally compare our system to familiar voice assistants, making it crucial for verbal responses to be both engaging and contextually relevant. Ensuring the quality of verbal interactions is vital to prevent premature conversation termination, which would hinder the study of nonverbal behaviors.

## VI. Conclusion

This paper introduces a modular, distributed architecture that advances SIAs by enabling them to engage in both verbal and nonverbal interaction in real-world settings. By combining advanced CAI, camera-based user management, and MetaHuman's realistic embodiment, our system facilitates the study of nonverbal behaviors. A notable strength is the robust integration of real-time facial animations with motion-captured gestures, all supported by a flexible dialog system that ensures locally compliant AI applications. The three-week field study in a museum highlighted both potentials for enhancing user interaction and challenges such as computational latency and environmental adaptation.

Ultimately, this architecture offers a robust and adaptable framework for exploring nonverbal behavior in dynamic real-world interaction studies, advancing the field of nonverbal behavior in SIAs. Future research should utilize this architecture to explore the nuanced effects of specific nonverbal behaviors, employing both generative and motion capture approaches, across diverse real-world settings. By deepening our understanding of the impact of nonverbal communication, particularly on user engagement and satisfaction, this work could significantly enhance the development of SIAs in diverse social contexts.

## Acknowledgment

The R&D activities were reviewed and approved by the Ethics Research Committee of TH Köln (application no. THK-2023-0004). The authors acknowledge the financial support by the Federal Ministry of Education and Research of Germany in the framework FH-Kooperativ 2-2019 (project number 13FH504KX9). We thank our collaboration partner, the Deutsches Museum Bonn, for their assistance and contributions.